\documentclass{llncs}

\usepackage{xcolor}
\usepackage{amsmath,amssymb,amsfonts}
\usepackage{mathtools}
\usepackage{xspace}
\usepackage{subfigure}
\usepackage{paralist}
\usepackage{microtype}
\usepackage{colonequals}
\usepackage{tikz}
\usepackage{cite}
\usepackage{multirow}
\usetikzlibrary{arrows.meta,decorations.pathmorphing,positioning,fit,trees,shapes,shadows,automata,calc,decorations.markings,patterns} 
\tikzset{>=latex}
\newcommand{\mc}[1]{\textcolor{green!50!black}{#1}}

\usepackage{algorithmicx,algorithm}
\usepackage[noend]{algpseudocode}
\usepackage{nicefrac}

\renewcommand{\emptyset}{\varnothing}

\usepackage{listings}

\definecolor{prismgreen}{HTML}{009900}
\definecolor{prismred}{HTML}{cc0000}
\definecolor{prismblue}{HTML}{0000cc}

\lstdefinelanguage{Prism}{
        basicstyle=\color{prismred}\scriptsize\ttfamily,
        literate=*	{0}{{\textcolor{prismblue}{0}}}{1}
			{1}{{\textcolor{prismblue}{1}}}{1}
			{2}{{\textcolor{prismblue}{2}}}{1}
			{3}{{\textcolor{prismblue}{3}}}{1}
			{4}{{\textcolor{prismblue}{4}}}{1}
			{5}{{\textcolor{prismblue}{5}}}{1}
			{6}{{\textcolor{prismblue}{6}}}{1}
			{7}{{\textcolor{prismblue}{7}}}{1}
			{8}{{\textcolor{prismblue}{8}}}{1}
			{9}{{\textcolor{prismblue}{9}}}{1}
			{.0}{{\textcolor{prismblue}{.0}}}{2}
			{.1}{{\textcolor{prismblue}{.1}}}{2}
			{.2}{{\textcolor{prismblue}{.2}}}{2}
			{.3}{{\textcolor{prismblue}{.3}}}{2}
			{.4}{{\textcolor{prismblue}{.4}}}{2}
			{.5}{{\textcolor{prismblue}{.5}}}{2}
			{.6}{{\textcolor{prismblue}{.6}}}{2}
			{.7}{{\textcolor{prismblue}{.7}}}{2}
			{.8}{{\textcolor{prismblue}{.8}}}{2}
			{.9}{{\textcolor{prismblue}{.9}}}{2}
			{[}{{\textcolor{black}{[}}}{1}
			{]}{{\textcolor{black}{]}}}{1}
			{+}{{\textcolor{black}{+}}}{1}
			{-}{{\textcolor{black}{-}}}{1}
			{=}{{\textcolor{black}{=}}}{1}
			{<}{{\textcolor{black}{<}}}{1}
			{>}{{\textcolor{black}{>}}}{1}
			{\&}{{\textcolor{black}{\&}}}{1}
			{|}{{\textcolor{black}{|}}}{1}
			{:}{{\textcolor{black}{:}}}{1}
			{;}{{\textcolor{black}{;}}}{1}
			{(}{{\textcolor{black}{(}}}{1}
			{)}{{\textcolor{black}{)}}}{1}
			{..}{{\textcolor{black}{..}}}{2},
        keywords= {bool,ceil,const,ctmc,double,dtmc,endinit,endmodule,endrewards, endsystem,F,false,floor,formula,G,global,I,init,int,label,max,mdp,min,module,nondeterministic,P,Pmin,Pmax,prob,probabilistic,rate,rewards,Rmin,Rmax,S,stochastic,system,true,U, option, either, assignment, relation, operation, hole, variable},
        keywordstyle={\bfseries\color{black}},
        numberstyle=\footnotesize\color{black},
        comment=[l] {//}, morecomment=[s]{/*}{*/},
        commentstyle= \color{prismgreen},
        tabsize=4,
        captionpos=b,
        escapechar=^,
        moredelim=[is][\color{orange}]{@}{@},
}
\newcommand{\init}{\ensuremath{s_0}}

\newcommand{\Succ}{\ensuremath{\mathsf{Succ}}}

\usepackage{algorithmicx,algorithm}
\usepackage[noend]{algpseudocode}

\newcommand{\familymc}{\ensuremath{\mathfrak{D}}}

\newcommand{\pathset}{\mathsf{Paths}}

\newcommand{\pathsfin}{\pathset_{\mathit{fin}}}

\newcommand{\act}{\ensuremath{a}}
\newcommand{\Act}{\ensuremath{\mathit{Act}}}
\newcommand{\last}[1]{\mathrm{last}(#1)}

\newcommand{\Distr}{\mathit{Distr}}

\newcommand{\distDom}{X}
\newcommand{\distFunc}{\mu}
\newcommand{\distDomElem}{x}

\newcommand{\dtmc}{\ensuremath{D}}
\newcommand{\mdp}{\ensuremath{M}}
\newcommand{\sched}{\ensuremath{\sigma}}
\newcommand{\induced}[2]{\ensuremath{{#1}_{#2}}}
\begin{document}

\title{Model Repair Revamped\thanks{This work has been supported by the DFG RTG 2236 ``UnRAVeL'', the ERC Advanced Grant 787914 ``FRAPPANT'', the Czech Science Foundation grant No. AUTODEV GA19-24397S, and the IT4Innovations excellence in science project No. LQ1602.}}
\subtitle{--- On the Automated Synthesis of Markov Chains ---}
\author{Milan \v{C}e\v{s}ka\inst{1}, Christian Dehnert\inst{2}, Nils Jansen\inst{3}, \\
Sebastian Junges\inst{2} and Joost-Pieter Katoen\inst{2}}
\institute{
Brno University of Technology, FIT, IT4I Centre of Excellence, Brno, Czech Republic
\and
RWTH Aachen University, Aachen, Germany 
\and 
Radboud University, Nijmegen, The Netherlands
}
\titlerunning{Revamping model repair}
\authorrunning{\v{C}e\v{s}ka, Dehnert, Jansen, Junges and Katoen}

\maketitle
\thispagestyle{plain}\pagestyle{plain}  %% restore page numbers

\begin{abstract}
	%!TEX root = main.tex

This paper outlines two approaches---based on counterexample-guided abstraction refinement (CEGAR) and counterexample-guided inductive synthesis (CEGIS), respectively---to the automated synthesis of finite-state probabilistic models and programs.
Our CEGAR approach iteratively partitions the design space starting from an abstraction of this space and refines this by a light-weight analysis of verification results.
The CEGIS technique exploits critical subsystems as counterexamples to prune all programs behaving incorrectly on that input.
We show the applicability of these synthesis techniques to sketching of probabilistic programs, controller synthesis of POMDPs, and software product lines.
\end{abstract}

\section{Introduction}
\paragraph{Model repair.}
In 2011, Smolka \emph{et al.}~\cite{bartocci2011model} coined the following \emph{model repair} problem~\cite{DBLP:journals/ai/BuccafurriEGL99}: given a finite Markov chain $\dtmc$ and a probabilistic specification $\varphi$ such that $\dtmc \not\models \varphi$, find a Markov chain $\dtmc'$ that differs from $\dtmc$ only in the transition probabilities, such that $\dtmc' \models \varphi$.
Typical probabilistic specifications impose a threshold on reachability probabilities, such as ``is the probability to reach a bad state at most $\nicefrac{1}{1000}$?''
Model repair thus amounts to tweaking (some of) the probabilities in a given Markov chain in order to obtain a chain satisfying the specification.
It can be solved using parameter synthesis~\cite{bartocci2011model} techniques as, e.g., supported by the Prophesy tool~\cite{dehnert2015prophesy}.
An extension of model repair in which repairs are associated a cost and a minimal-total-cost repair is to be found can be solved by non-linear programming~\cite{bartocci2011model}. 
The scalability of model repair can be improved in several ways: by solving a series of convex programs instead of a non-linear program~\cite{DBLP:conf/atva/CubuktepeJJKT18}, by repairing abstractions of Markov chains~\cite{DBLP:journals/corr/Chatzieleftheriou15}, or by a greedy approach exploiting monotonicity~\cite{pathak-et-al-nfm-2015}.
Particle swarm optimisation has been used to model repair of Markov decision processes (MDPs)~\cite{DBLP:conf/tase/ChenHHKQ013}. 

\paragraph{Topology changes.}
In the original setting of model repair, only transition probabilities are subject to change.
Adding or removing transitions is not admitted.
That is to say, every possible repair keeps the topology of the Markov chain invariant.
This notion of repair is thus in fact \emph{transition probability repair}.
In this paper, we take this idea a step further, and allow for amending the Markov chain's topology.
More precisely, in addition to the possibility of modifying transition probabilities, we consider the possible deletion and/or addition of transitions. 
Changing the topology results in varying sets of reachable states.
Viewing Markov chains as operational model for (discrete) probabilistic programs, repairs may affect the control structure as well as the probabilistic choices.
Traditional parameter synthesis techniques are inadequate for this problem, as these techniques restrict parameter expressions like $\nicefrac{1}{2}{-}\varepsilon$ to range over $(0,1)$, i.e., excluding zero (no transition) and one (a Dirac transition).
Topology changes often come at a price; e.g., for parametric Markov chains the complexity of qualitative (i.e., zero-one) reachability becomes NP-complete whereas in absence of topology changes a polynomial-time algorithm suffices~\cite{DBLP:journals/corr/Chonev17}.

\paragraph{Synthesis.}
Model (or: program) repair in the aforementioned sense is a natural instance of model (or: program) synthesis~\cite{DBLP:journals/cacm/AlurSFS18,DBLP:conf/kbse/GerasimouTC15,DBLP:journals/ftpl/GulwaniPS17}.
Program synthesis amounts to automatically provide an instantiated probabilistic program satisfying all quantitative properties, or returns that such design does not exist. 
Though the synthesis problem is undecidable in general, there are interesting sub-cases --- such as our variant of model repair --- that are decidable (but computationally intensive).
We consider a family $\familymc$ of Markov chains where each family member can be viewed as an admissible repaired version of Markov chain $D \in \familymc$.
Families are finite and consist of finite chains.
As in syntax-guided synthesis, possible repairs are described by some grammar rules, either at the model level or in some high-level description language such as the probabilistic guarded command language~\cite{DBLP:journals/toplas/MorganMS96} or the PRISM modelling language~\cite{KNP11}.
The successful approach of program sketching~\cite{DBLP:journals/sttt/Solar-Lezama13} naturally fits within this setting.
Starting from a program sketch, i.e., a program with ``holes'', it aims to obtain a program satisfying the specification $\varphi$ by filling the holes with possible repairs.
Holes are the unknown parts of the program and can be replaced by one of finitely many options.
All possible program sketch realisations constitute the family $\familymc$.
The synthesis problems considered in this paper can be used to answer queries like: (a) give a possible repair (if one exists)?, (b) what are all possible repairs?, and (c) which repairs are optimal in the sense of minimising the probability to reach a bad state?
Each of these queries can be considered under the additional constraint of minimising the total cost of repairs, e.g., what are all possible repairs for $\varphi$ that are cost minimal?

\paragraph{Our synthesis approaches.}
A naive enumerative solution is to analyse each individual family member, i.e., each possible amendment, or each possible hole instantiation.
This is infeasible for large families.
We therefore outline two approaches to the automated synthesis of finite-state probabilistic models: the first one fits within the realm of counterexam\-ple-guided abstraction refinement (CEGAR~\cite{DBLP:conf/cav/ClarkeGJLV00}) while the second approach fits within counterexample-guided inductive synthesis (CEGIS)~\cite{DBLP:conf/pldi/Solar-LezamaRBE05}.
Full details of the approaches can be found in~\cite{DBLP:conf/tacas/CeskaJJK19,CeskaDJK19}.
We present both techniques at the level of Markov chains for threshold problems on reachability problems and illustrate their usage on simple probabilistic programs.

\begin{figure}[t]
\tikzset{
        hatch distance/.store in=\hatchdistance,
        hatch distance=6pt,
        hatch thickness/.store in=\hatchthickness,
        hatch thickness=1pt
    }
    \makeatletter
    \pgfdeclarepatternformonly[\hatchdistance,\hatchthickness]{flexible hatch}
    {\pgfqpoint{0pt}{0pt}}
    {\pgfqpoint{\hatchdistance}{\hatchdistance}}
    {\pgfpoint{\hatchdistance-1pt}{\hatchdistance-1pt}}%
    {
        \pgfsetcolor{\tikz@pattern@color}
        \pgfsetlinewidth{\hatchthickness}
        \pgfpathmoveto{\pgfqpoint{0pt}{0pt}}
        \pgfpathlineto{\pgfqpoint{\hatchdistance}{\hatchdistance}}
        \pgfusepath{stroke}
    }
    \makeatother

\centering
\subfigure[The CEGAR approach]{
    \scalebox{0.8}{
    	\begin{tikzpicture}
    	
      \newcommand{\tikzsmallerfontsize}{\footnotesize}
    		\node[] (in) {family};
    		\node[minimum width=10em, minimum height=0.5em, draw, inner sep=3pt,below=0.24cm of in] (abstract) {\tikzsmallerfontsize abstract to quotient};
    		
    		\node (quots) [cylinder, shape border rotate=0, draw, minimum height=10mm, minimum width=7.5mm, below=0.3cm of abstract, xshift=-0.5cm] {\tikzsmallerfontsize quotients};
    		
    		\node (mc) [draw, rectangle, right=1.2cm of quots, minimum width=8em, inner sep=3pt] {\tikzsmallerfontsize verify $M \models \varphi$};
    		\node (spurious) [draw, minimum width=20mm, align=center, diamond, below=0.7cm of mc, aspect=2.8] {\tikzsmallerfontsize satisfied?};
    		\node (split) [draw, rectangle, left=1.1cm of spurious, minimum width=8em, inner sep=3pt, align=center] {\tikzsmallerfontsize refine quotient:\\split family};
    		\node (res) [below=0.3cm of spurious] {feasible realisation};
    		\node (nores) [left=2.3cm of res] {unsatisfiable};
    		\coordinate (a) at (abstract.240);
    		
    		\draw[->] (abstract.south) -| node [pos=0.5,right] {\tikzsmallerfontsize } (quots.north);
    		\draw[->] (quots) edge node [pos=0.5,below] {\tikzsmallerfontsize pick $M$} (mc);
    		\draw[->] (mc) edge node [pos=0.5, align=left, xshift=-0.5em] {\tikzsmallerfontsize $\!\!\Pr(M \models \varphi)$ (+ CE)} (spurious);
    		\draw[->] (spurious) edge node [pos=0.4,right] {\tikzsmallerfontsize yes} (res);
    		\draw[->] (spurious) edge node [pos=0.2,below] {\tikzsmallerfontsize inconclusive} (split);
    		\draw[-] (spurious.east) edge node [pos=0.2,below] {\tikzsmallerfontsize no}  ($(spurious.east)+(0.4,0)$);
    		
	    	\draw[-] ($(spurious.east)+(0.4,0)$) -- ($(spurious.east)+(0.4,1.7)$);
	    	\draw[->]  ($(spurious.east)+(0.4,1.7)$) edge[bend right=16] (quots.20);
    		
    		\draw[->] (split) edge node [pos=0.5,right, align=left] {\tikzsmallerfontsize } (quots);
    		\draw[->] (in) -- (abstract);
    		\draw[->] (quots) -| node[above] {empty} (nores.160);
    		
    	\end{tikzpicture}
    	}
    \label{fig:CEGAR}
    }
      \subfigure[The CEGIS approach]{

    \scalebox{0.8}{
      \begin{tikzpicture}
      \tikzset{inv/.style={inner sep=0pt, minimum size=0pt}}
      \newcommand{\tikzEdgeLength}{1.2cm}
      \newcommand{\tikzsmallerfontsize}{\footnotesize}
    
      \node[minimum width=\tikzEdgeLength, minimum height=\tikzEdgeLength, draw] (synthesiser) {\phantom{Synthesiser}};
          
      \node[right=2*\tikzEdgeLength of synthesiser.east, minimum width=\tikzEdgeLength, minimum height=\tikzEdgeLength, draw, minimum width=\tikzEdgeLength, rounded corners] (verifier) {\small Verifier};
      
      \path[pattern color=lightgray,pattern=flexible hatch] ($(synthesiser.north east)-(0.01,0.01)$) -- ($(synthesiser.north east)-(0.4,0.01)$) to[draw,  bend right] ($(synthesiser.south)+(0.15,0.01)$) -- ($(synthesiser.south east)+(-0.01,0.01)$) -- cycle;
      \draw ($(synthesiser.north east)-(0.4,0)$) to[draw,  bend right] ($(synthesiser.south)+(0.15,0.01)$);
      
      \node[inv, circle, inner sep=0.7pt, fill=black] (instance) at ($(synthesiser.north east)-(0.3,0.5)$) {};
      
      \draw[->, shorten <= 1pt, shorten >= 1pt] (instance) (instance) to[bend left] node[auto, pos=0.6, yshift=-1pt] {\tikzsmallerfontsize instance} (verifier);
      \draw[->] (verifier) to[bend left] node[auto, swap, near start, xshift=5pt, yshift=-3pt] {\tikzsmallerfontsize reject +} node[auto, xshift=10pt, yshift=2pt] (cex) {\tikzsmallerfontsize CE} ($(instance)-(0.2,0.4)$);
      
      \coordinate (topleft) at ($(synthesiser.north west)+(-0.15,0.4)$);
      \coordinate (bottomright) at ($(verifier.south east)+(0.15,-0.55)$);

      \draw[densely dotted, rounded corners] (topleft) rectangle (bottomright);
      
      \node[above=0.55cm of synthesiser] (sketch) {\tikzsmallerfontsize family};
      \draw[->] (sketch) to (synthesiser);

      \node (properties) at (verifier |- sketch) {\tikzsmallerfontsize properties};
      \draw[->] (properties) to (verifier);

      \node[below=0.75cm of synthesiser] (unsat) {\tikzsmallerfontsize unsatisfiable};
      \draw[->] (synthesiser) to node[near start, xshift=11pt] {\tikzsmallerfontsize no instance} (unsat);

      \node[align=center, anchor=north] (synthprog) at (verifier |- unsat.north) {\tikzsmallerfontsize feasible realisation};
      \draw[->] (verifier) to node[left] {\tikzsmallerfontsize accept} (synthprog);
      
    \end{tikzpicture}
    
     \label{fig:CEGIS}
    }
    }
        
  \caption{CEGAR and CEGIS approaches to (feasibility) synthesis.}
 
\end{figure}
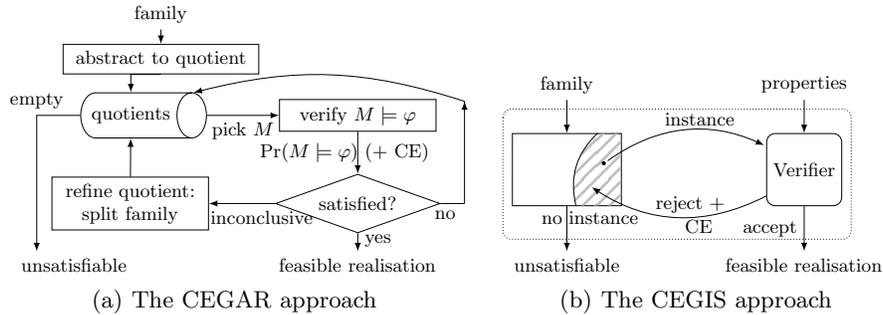

\paragraph{Using CEGAR.}
We represent all possible designs, thus the entire family, by a single MDP.
A single, initial, non-deterministic choice determines according to which family member the MDP behaves.
This technique is adopted from~\cite{DBLP:journals/fac/ChrszonDKB18} and originated in software product lines~\cite{DBLP:journals/sttt/ClassenCHLS12}.
As the MDP can be prohibitively large, we do not solve the synthesis problems directly on this model, but rather on an abstraction of it --- similar in spirit as repairing abstract models~\cite{DBLP:journals/corr/Chatzieleftheriou15}.
Verifying the abstraction, i.e., the quotient MDP $M$, yields under- and over-approximations of the min and max probability of satisfying $\varphi$, respectively.
A repair is impossible if e.g., the verification reveals that the min probability exceeds $r$ for $\varphi$ with threshold $\leq r$.
If the model checking is inconclusive, i.e., the abstraction is too coarse, we iteratively
refine the quotient MDP by splitting the family into sub-families, see Fig.~\ref{fig:CEGAR}.
The refinement is guided using the so-called inconsistent schedulers (aka: counterexamples) that optimise the probability on the MDP. 

%%%

\paragraph{Using CEGIS.}
Starting from a family, a candidate realisation $D$ is selected and discharged to a verifier, see Fig.~\ref{fig:CEGIS}.
Using off-the-shelf probabilistic model-checking techniques~\cite{DBLP:conf/lics/Katoen16,DBLP:reference/mc/BaierAFK18}, it verifies whether $D \models \varphi$ in case a solution is found.
If $D \not\models \varphi$, a counterexample (CE) is derived which in our setting is a fragment~\cite{DBLP:journals/tcs/WimmerJAKB14} of the realisation $D$ that refutes $\varphi$.
The key is that this CE is exploited in a clever way by an SMT (satisfiability modulo theory)-based synthesiser to rule out potentially many realisations (the dashed area in Fig.~\ref{fig:CEGIS}) --- rather than the just refuted realisation $D$ --- at once. 
Thus, in a sense counterexamples are ``extended'' to a set of refuting realisations.
This synthesis-verification loop is repeated until either a satisfying realisation is found or the entire family is pruned implying the non-existence of a realisation $D \models \varphi$.

\paragraph{Design space partitioning.}
Both the CEGAR and CEGIS approach iteratively partition the family into ``good'', ``bad'' and inconclusive realisations.
Phrased in terms of model repair, they partition the family into repaired, failed, and unknown Markov chains.
The two approaches use complementary partitioning strategies.
Whereas the CEGAR approach starts from considering all possible realisations, and successively splits the entire family of realisations into sub-families, the CEGIS approach starts with a single candidate realisation, and rules out several realisations by effectively exploiting counterexamples.

\section{Preliminaries}
We start with basic foundations, for details, see~\cite{BK08,DBLP:reference/mc/BaierAFK18}.
Then, we formalise the notion of families of Markov chains, 
and define various synthesis problems.

\subsection{Probabilistic Models and Specifications}

\paragraph{Probabilistic models.}
A \emph{probability distribution} over countable set $\distDom$
is a function $\distFunc\colon\distDom\rightarrow [0,1]$ with $\sum_{\distDomElem\in\distDom}\distFunc(\distDomElem)=\distFunc(\distDom)=1$.
Let $\Distr(\distDom)$ denote the set of all distributions on $\distDom$. 
%% The support of a distribution $\distFunc$ is $\supp(\distFunc) = \{x\in\distDom \mid \distFunc(x)>0\}$.
%% A distribution is \emph{Dirac} if $|\!\supp(\distFunc)| = 1$.

\begin{definition}[MC]\label{def:dtmc}
A \emph{Markov chain} (MC) $\dtmc = (S,\init, \mathbf{P})$ with finite set $S$ of states, initial state $\init \in S$, and transition probability function
$\mathbf{P}\colon S \rightarrow \Distr(S)$.
%% is a transition probability matrix.
% (i.e~$\forall s\in S\colon \sum_{s'\in S} \mathbf{P}(s,s') = 1$), 
\end{definition}
MCs have unique distributions over successor states at each state. 
A sub-Markov chain (sub-MC) is induced by a MC and a subset of its states.
For $X \subseteq S$, let the set $\Succ(X)$ denote the successor states of $X$, i.e., $\Succ(X) = \{ t \in S \mid \exists s \in X.~\mathbf{P}(s,t) > 0\}$.
\begin{definition}[sub-MC]
Let MC $D = (S,\init, \mathbf{P})$ and $C \subseteq S$ a set of \emph{(critical) states} with $\init \in C$.
The \emph{sub-MC} of $D,C$ is the MC $D' = (S', \init, \mathbf{P}')$ with
$S' = C \cup \Succ(C)$, and \[\mathbf{P}'(s,t) = \begin{cases}
 \mathbf{P}(s,t) & s \in C, t \in S \\
 1  & s \in \Succ(C) \setminus C \land t = s \\
 0 & \text{otherwise.}
 \end{cases}
 \]
\end{definition}
Adding non-determinism over distributions leads to Markov decision processes.
\begin{definition}[MDP]\label{def:mdp}	
A \emph{Markov decision process} (MDP) is a tuple $\mdp=(S,\init,\Act,\mathcal{P})$ where $S, \init$ as in Def.~\ref{def:dtmc}, $\Act$ is a finite set of actions, and 
$\mathcal{P}\colon S\times \Act \nrightarrow \Distr(S)$ is a partial transition probability function.
\end{definition}

\noindent
The \emph{available actions} in $s\in S$ are $\Act(s)=\{\act\in\Act\mid \mathcal{P}(s,\act) \neq \bot \}$. An MDP with $|\Act(s)|=1$ for all $s\in S$ is an MC. 
%% For MCs (and MDPs), a state-reward function is $\textit{rew}\colon S\rightarrow \mathbb{R}_{\geq 0}$.
%% The reward $\textit{rew}(s)$ is earned upon leaving~$s$.
A \emph{path} of an MDP $M$ is an (in)finite sequence $\pi = s_0\xrightarrow{\act_0}s_1\xrightarrow{\act_1}\cdots$,
where $s_i\in S$, $\act_i\in\Act(s_i)$, and $\mathcal{P}(s_i,\act_i)(s_{i+1})\neq 0$ for all $i\in\mathbb{N}$.
For finite $\pi$, $\last{\pi}$ denotes
the last state of $\pi$.
Let $\pathsfin^{M}$ denote the set of finite paths of $M$.
The notions of paths carry over to MCs (actions are omitted).

%\paragraph{Schedulers.} 
%We use standard probability and expected reward measures over MCs.
%
\begin{definition}[Scheduler]	\label{def:scheduler}
A \emph{scheduler} for an MDP $\mdp=(S,\init,\mathit{Act},\mathcal{P})$ is a function $\sched\colon\pathsfin^{M}\to\Act$
  such that $\sched(\pi)\in \Act(\last{\pi})$ for all $\pi\in \pathsfin^{\mdp}$.
   Scheduler $\sched$ is \emph{memoryless} if $\last{\pi}=\last{\pi'} \implies \sched(\pi)=\sched(\pi')$ for all $\pi,\pi'\in\pathsfin^{\mdp}$.
  Let $\Sigma^{\mdp}$ denote the set of all schedulers of $\mdp\!$.
\end{definition}
Schedulers resolve the non-determinism over actions in the MDP.
Applying scheduler $\sched$ to an MDP $\mdp$ yields the \emph{induced} Markov chain $\induced{\mdp}{\sched}$.
\paragraph{Specifications.} 
For simplicity, we only consider reachability specifications $\varphi=\mathbb{P}_{\sim \lambda}(\lozenge G)$ where $G \subseteq S$ is a set of goal states, $\lambda\in [0,1]\subseteq{\mathbb{R}}$ is a threshold, and ${{}\sim{}} \in \{<,\leq,\geq,>\}$ is a binary comparison operator.
Extensions to expected rewards, PCTL*~\cite{aziz1995usually}, or $\omega$-regular properties are rather straightforward.

The interpretation of $\varphi$ for MC $\dtmc$ is as follows.
Let $\texttt{Prob}(\dtmc,\phi)(s)$ denote the probability to satisfy $\phi =\lozenge G$ from state $s\in S$ in MC $\dtmc$.
For initial state $\init$, we abbreviate $\texttt{Prob}(\dtmc,\phi)(\init)$ by $\texttt{Prob}(\dtmc,\phi)$.
Then, $\dtmc\models\varphi$ iff $\texttt{Prob}(\dtmc, \phi) \sim \lambda$.
%% Analogously, we define expected reward specifications of the form $\varphi=\mathbb{E}_{\sim \kappa}(\lozenge G)$ with $\kappa\in {\mathbb{R}_{\geq 0}}$. %% We refer to $\lambda$/$\kappa$ as \emph{thresholds}.
The specification $\varphi$ holds in MDP $M$, denoted $M\models\varphi$, iff it holds for the induced MCs under all schedulers.
The maximum probability 
%% $\texttt{Prob}^{\max}(\mdp,\phi)$ 
to satisfy $\phi$ in MDP $M$ is given by a maximising scheduler $\sigma^*\in\Sigma^M$, i.e., there is no scheduler $\sigma'\in\Sigma^M$ with $\texttt{Prob}(\induced{\mdp}{\sigma^*},\phi)<\texttt{Prob}(\induced{\mdp}{\sigma'},\phi)$.
As we consider finite models, such a maximising scheduler always exists. 
Minimum probabilities are defined analogously.
%% Analogously, we define the minimising probability $\texttt{Prob}^{\min}(\mdp,\phi)$, and the maximising (minimising) expected reward $\texttt{ExpRew}^{\max}(\mdp, \phi)$ ($\texttt{ExpRew}^{\min}(\mdp, \phi)$).
%% ($\texttt{ExpRew}(\dtmc,\phi)(s)$).
%% The notation is analogous for maximising and minimising probability 
%% and expected reward 
%% measures in MDPs.
%
%\begin{remark}[Well-defined expected reward]\label{rem:well-defined-exp-rew}
%Note that the expected reward $\texttt{ExpRew}(\dtmc, \phi)$ to satisfy path formula $\phi$ is only defined if $\texttt{Prob}(\dtmc, \phi)=1$.
%Accordingly, the expected reward for MDP $\mdp$ under scheduler $\sigma\in\Sigma^\mdp$ requires $\texttt{Prob}(\mdp_{\sigma}, \phi)=1$.	
%\end{remark}

%%% JPK: rewards have been commented out. 

\subsection{Families of Markov Chains}

We present our approaches on an explicit representation of a \emph{family of MCs} using a parametric transition probability function.
Such an explicit representation allows to reason about practically interesting synthesis problems, see Sect.~\ref{sec:applications}.
\begin{definition}[Family of MCs]
A \emph{family of MCs} is a tuple $\familymc =(S,\init,K,\mathfrak{P})$ where $S$ and $\init$ are as before, $K$ is a finite set of discrete parameters such that the domain of each parameter $k\in K$ is $T_k\subseteq S$, and $\mathfrak{P}\colon S \rightarrow \Distr(K)$. %% is a family of transition probability matrices.
\end{definition}
The transition probability function of MCs maps states to distributions over successor states. 
For families of MCs, this function maps states to distributions over parameters.
Instantiating each of these parameters with a value from its domain yields a ``concrete'' MC, called a \emph{realisation}.

\begin{definition}[Realisation]
	A \emph{realisation} of a family $\familymc=(S,\init,K,\mathfrak{P})$ is a function $r\colon K \rightarrow S$ where $\forall k\in K\colon r(k) \in T_k$. 
	A realisation $r$ yields an MC $D_r = (S,\init,\mathfrak{P}(r))$, where $\mathfrak{P}(r)$ is the transition probability matrix in which 
	each  
	$k\in K$ in $\mathfrak{P}$ 
	is replaced by $r(k)$.
	Let $\mathcal{R}^\familymc$ denote the \emph{set of all realisations} for $\familymc$.
\end{definition}
As a family $\familymc$ of MCs is defined over finite parameter domains, the number of family members (i.e. realisations from $\mathcal{R}^\familymc$) of $\familymc$ is finite, viz. $|\familymc| \colonequals |\mathcal{R}^\familymc|  = \prod_{k\in K}|T_k|$, but exponential in $|K|$.
Subsets of $\mathcal{R}^\familymc$ induce so-called \emph{subfamilies} of $\familymc$.
While all these MCs share the same state space, their \emph{reachable} states may differ, as demonstrated by the following example.

\begin{example}[Family of MCs]\label{ex:dtmc_family}
	Consider the family of MCs $\familymc = (S, \init, K, \mathfrak{P})$
	where $S=\{0,\hdots,4\}$, $\init = 0$, and $K=\{k_0, \hdots, k_5\}$ with domains 
	$T_{k_0} = \{ 0 \}$, $T_{k_1} = \{ 1 \}$, $T_{k_2} = \{ 2, 3 \}$, $T_{k_3} = \{ 2, 4 \}$, $T_{k_4} = \{ 3 \}$ and $T_{k_5} = \{ 4 \}$.
	The parametric transition function is defined by:
	\begin{align*}
	&	\mathfrak{P}(0) = 0.5 \colon k_1 + 0.5 \colon k_2 & 
	& \mathfrak{P}(1) = 0.1 \colon k_0 + 0.9 \colon k_1 & 
   &  \mathfrak{P}(2) = 1\colon k_3 \\ 
 & \mathfrak{P}(3) = 0.8 \colon k_3 + 0.2 \colon k_4 & 
&	\mathfrak{P}(4) = 1\colon k_5 &	
	\end{align*}
We can simplify the representation by substituting the constants:
	\begin{align*}
	&	\mathfrak{P}(0) = 0.5 \colon 1 + 0.5 \colon k_2 & 
	& \mathfrak{P}(1) = 0.1 \colon 0 + 0.9 \colon 1 & 
   &  \mathfrak{P}(2) = 1\colon k_3 \\ 
 & \mathfrak{P}(3) = 0.8 \colon k_3 + 0.2 \colon 3 & 
&	\mathfrak{P}(4) = 1\colon 4 &	
	\end{align*}
Fig.~\ref{fig:realisation} shows the four MCs that result from the realisations $\{r_1,r_2,r_3,r_4\}=\mathcal{R}^\familymc$ of $\familymc$.
% for the parameters $k_1$ and $k_2$.
States that are unreachable from the initial state are greyed out.
The family has five states, each of which are reachable in one of the realisations. Yet, every realisation has at most four reachable states.
\end{example}
\begin{figure}[t]
	\centering
	\subfigure[$D_{r_1}$ with $r_1(k_2)=2, r_1(k_3)=2$]{%
	    \label{fig:realisation_0_2}
		\scalebox{0.7}{\begin{tikzpicture}[every node/.style={circle}]
	\node[draw] (1) {$0$};
	\node[draw, right=1.2 cm of 1] (0) {$1$} ;
	\node[draw, right=1.2 cm of 0] (2) {$2$};
	\node[draw=gray, right=1.2 cm of 2] (3) {\color{gray}$3$};
	\node[draw=gray, right=1.2 cm of 3] (4) {\color{gray}$4$};

	\draw[->] (0) edge[loop above] node[auto] {$0.9$} (0);
	\draw[->] (0) edge[bend right=60] node[above] {$0.1$} (1);
	\draw[->] (1) edge[] node[above] {$0.5$} (0);
	\draw[->] (1) edge[bend right=20] node[pos=0.1,below] {$0.5$} (2);
	\draw[->] (2) edge[loop above] node[auto] {$1$} (2);

	\draw[->,gray] (4) edge[loop above] node[auto] {$1$} (4);
	\draw[->,gray] (3) edge[] node[above] {$0.8$} (2);
	%\draw[->,gray] (1) edge[bend right=20] node[pos=0.1,below] {$0.5$} (3);
	
	%\draw[->,gray] (3) edge[] node[below] {$0.8$} (4);
	%\draw[->,gray] (2) edge[bend left=20] node[above, near start] {$1$} (4);

	\draw[->,gray] (3) edge[loop below] node[left] {$0.2$} (3);

%	\draw[->,gray] (1) edge[] node[below] {$0.5$} (2);
%	\draw[->] (0) edge[loop below] node[auto] {$0.5$} (0);
%	\draw[->,gray] (2) edge[loop below] node[auto] {$1$} (2);
%	\draw[->,gray] (3) edge[bend right=20] node[above, near start] {$0.5$} (0);
%	\draw[->,gray] (3) edge[] node[below] {$0.5$} (2);
	\draw ($(1.north) + (0,0.3)$) edge[->] (1);
	\draw  [use as bounding box, draw=white] (-0.2,-1.4) rectangle (8.1,1.3) {};%different BB to other pics because of weird spacing
%different BB to other pics because of weird spacing
\end{tikzpicture}%
}}
	\subfigure[$D_{r_2}$ with $r_2(k_2)=2, r_2(k_3)=4$]{
	    \label{fig:realisation_1_2}
		\scalebox{0.7}{\begin{tikzpicture}[every node/.style={circle}]
	\node[draw] (1) {$0$};
	\node[draw, right=1.2 cm of 1] (0) {$1$} ;
	\node[draw, right=1.2 cm of 0] (2) {$2$};

	\node[draw=gray, right=1.2 cm of 2] (3) {\color{gray}$3$};
	\node[draw, right=1.2 cm of 3] (4) {$4$};

	\draw[->] (0) edge[loop above] node[auto] {$0.9$} (0);
	\draw[->] (0) edge[bend right=60] node[above] {$0.1$} (1);
	\draw[->] (1) edge[] node[above] {$0.5$} (0);
	\draw[->] (1) edge[bend right=20] node[pos=0.1,below] {$0.5$} (2);
	%\draw[->] (2) edge[loop above] node[auto] {$1$} (2);

	\draw[->] (4) edge[loop above] node[auto] {$1$} (4);
	%\draw[->,gray] (3) edge[] node[above] {$0.8$} (2);
	%\draw[->,gray] (1) edge[bend right=20] node[pos=0.1,below] {$0.5$} (3);
	
	\draw[->,gray] (3) edge[] node[below] {$0.8$} (4);
	\draw[->] (2) edge[bend left=20] node[above, near start] {$1$} (4);

	\draw[->,gray] (3) edge[loop below] node[left] {$0.2$} (3);

%	\draw[->,gray] (1) edge[] node[below] {$0.5$} (2);
%	\draw[->] (0) edge[loop below] node[auto] {$0.5$} (0);
%	\draw[->,gray] (2) edge[loop below] node[auto] {$1$} (2);
%	\draw[->,gray] (3) edge[bend right=20] node[above, near start] {$0.5$} (0);
%	\draw[->,gray] (3) edge[] node[below] {$0.5$} (2);
	\draw ($(1.north) + (0,0.3)$) edge[->] (1);
	\draw  [use as bounding box, draw=white] (-1,-1.4) rectangle (7.2,1.3) {};%different BB to other pics because of weird spacing
%different BB to other pics because of weird spacing
%different BB to other pics because of weird spacing
\end{tikzpicture}%
}}
	\subfigure[$D_{r_3}$ with $r_3(k_2)=3, r_3(k_3)=2$]{%
	    \label{fig:realisation_1_3}
		\scalebox{0.7}{\begin{tikzpicture}[every node/.style={circle}]
	\node[draw] (1) {$0$};
	\node[draw, right=1.2 cm of 1] (0) {$1$} ;
	\node[draw, right=1.2 cm of 0] (2) {$2$};

	\node[draw, right=1.2 cm of 2] (3) {$3$};
	\node[draw=gray, right=1.2 cm of 3] (4) {\color{gray}$4$};

	\draw[->] (0) edge[loop above] node[auto] {$0.9$} (0);
	
	\draw[->] (0) edge[bend right=60] node[above] {$0.1$} (1);
	\draw[->] (1) edge[] node[above] {$0.5$} (0);
	%\draw[->,gray] (1) edge[] node[auto] {$0.5$} (2);
	\draw[->] (2) edge[loop above] node[auto] {$1$} (2);

	\draw[->, gray] (4) edge[loop above] node[auto] {$1$} (4);
	\draw[->] (3) edge[] node[above] {$0.8$} (2);
	\draw[->] (1) edge[bend right=20] node[pos=0.1,below] {$0.5$} (3);
	
	%\draw[->,gray] (3) edge[] node[below] {$0.8$} (4);
	%\draw[->,gray] (2) edge[bend left=20] node[above, near start] {$1$} (4);

	\draw[->] (3) edge[loop below] node[left] {$0.2$} (3);

%	\draw[->,gray] (1) edge[] node[below] {$0.5$} (2);
%	\draw[->] (0) edge[loop below] node[auto] {$0.5$} (0);
%	\draw[->,gray] (2) edge[loop below] node[auto] {$1$} (2);
%	\draw[->,gray] (3) edge[bend right=20] node[above, near start] {$0.5$} (0);
%	\draw[->,gray] (3) edge[] node[below] {$0.5$} (2);
	\draw ($(1.north) + (0,0.3)$) edge[->] (1);
	\draw  [use as bounding box, draw=white] (-0.2,-1.4) rectangle (8.1,1.3) {};%different BB to other pics because of weird spacing
\end{tikzpicture}%
}}
	\subfigure[$D_{r_4}$ with $r_4(k_2)=3, r_4(k_3)=4$]{%
	    \label{fig:realisation_0_3}
		\scalebox{0.7}{\begin{tikzpicture}[every node/.style={circle}]
	\node[draw] (1) {$0$};
	\node[draw, right=1.2 cm of 1] (0) {$1$} ;
	\node[draw=gray,right=1.2 cm of 0] (2) {\color{gray}$2$};
	\node[draw, right=1.2 cm of 2] (3) {$3$};
	\node[draw, right=1.2 cm of 3] (4) {$4$};

	\draw[->] (0) edge[loop above] node[auto] {$0.9$} (0);
	
	\draw[->] (0) edge[bend right=60] node[above] {$0.1$} (1);
	\draw[->] (1) edge[] node[above] {$0.5$} (0);
	%\draw[->,gray] (1) edge[] node[auto] {$0.5$} (2);
	%\draw[->] (2) edge[loop above] node[auto] {$1$} (2);

	\draw[->] (4) edge[loop above] node[auto] {$1$} (4);
	%\draw[->,gray] (3) edge[] node[above] {$0.8$} (2);
	\draw[->] (1) edge[bend right=20] node[pos=0.1,below] {$0.5$} (3);
	
	\draw[->] (3) edge[] node[below] {$0.8$} (4);
	\draw[->,gray] (2) edge[bend left=20] node[above, near start] {$1$} (4);

	\draw[->] (3) edge[loop below] node[left] {$0.2$} (3);

%	\draw[->,gray] (1) edge[] node[below] {$0.5$} (2);
%	\draw[->] (0) edge[loop below] node[auto] {$0.5$} (0);
%	\draw[->,gray] (2) edge[loop below] node[auto] {$1$} (2);
%	\draw[->,gray] (3) edge[bend right=20] node[above, near start] {$0.5$} (0);
%	\draw[->,gray] (3) edge[] node[below] {$0.5$} (2);
	\draw ($(1.north) + (0,0.3)$) edge[->] (1);
	\draw  [use as bounding box, draw=white] (-1,-1.4) rectangle (7.2,1.3) {};%different BB to other pics because of weird spacing
\end{tikzpicture}%
}}
		\vspace{-0.5em}
		\caption{The four different realisations of family $\familymc$.}
		
		\label{fig:realisation}
\end{figure}

\subsection{Synthesis Problems}

\begin{problem}[Synthesis]
Let $\familymc$ be a family of MCs and $\varphi =\mathbb{P}_{\sim \lambda}(\phi)$ with $\phi = \lozenge G$ for $G\subseteq S$. We consider the following synthesis problems:
\begin{enumerate}
\item Find a realisation $r \in \mathcal{R}^{\familymc}$ with $D_r \models \varphi$.
\item Partition $\mathcal{R}^{\familymc}$ into $T$ and $F$ with $r \in T$ iff $D_r \models \varphi$ and $r \in F$ otherwise.
\item Find a realisation $r^* \in \mathcal{R}^{\familymc}$ with $r^* = \arg\!\!\max\limits_{r\in \mathcal{R}^{\familymc}} \{\texttt{Prob}(D_{r}, \phi)\}$.
\end{enumerate}
\end{problem}

\noindent
The first synthesis problem (referred to as feasibility synthesis) is to determine a realisation satisfying $\varphi$, provided some exists.
The second problem (referred to as threshold synthesis) is to identify the set of realisations satisfying and violating a given specification, respectively.
The feasibility synthesis problem is in a sense just a simple instance of threshold synthesis to find one realisation $r \in T$.
The last problem (referred to as max synthesis) is to find a realisation that maximises the reachability probability.
It can be defined for minimising such probabilities in a similar way.
As our families are finite, such optimal realisations $r^*$ always exist.
Phrased in terms of model repair, the first problem is concerned with the question whether a possible repair (under all admissible repairs) does exist, the second problem partitions the realisations into those that are repaired and those that cannot, while the last problem  is about finding the repair that maximises (or, dually, minimises) the objective.
The simplest synthesis problem, feasibility, is NP-complete~\cite{DBLP:conf/tacas/CeskaJJK19}.

%We state two synthesis problems for families of MCs. 
%The first is to identify the set of MCs satisfying and violating a given specification, respectively.
%The second is to find a MC that maximises/minimises a given objective. 
%We call these two problems \emph{threshold synthesis} and \emph{max/min synthesis}. 
%
%\begin{problem}[Threshold synthesis]\label{prob:threshold} 
%Let $\familymc$ be a family of MCs and $\varphi$ a probabilistic reachability or expected reward specification.
%%, and $\mathcal{R}^\familymc$ the set of all realisations of $\familymc$.
%The \emph{threshold synthesis problem} is to partition $\mathcal{R}^{\familymc}$ into %two sets of subspaces
% $T$ and $F$ such that $\forall r \in T\colon D_r \vDash \varphi$ and $\forall r \in F\colon D_r \nvDash \varphi$.% and $\bigcup_{\mathcal{R} \in T \cup F} \mathcal{R} = \mathcal{R}^{\familymc}$.
%\end{problem}
%%
%As a special case of the threshold synthesis problem, the \emph{feasibility synthesis problem} is to find just one realisation $r\in \mathcal{R}^{\familymc}$ such that $D_r \vDash \varphi$.
%%
%\begin{problem}[Max synthesis]\label{prob:max} 
%Let $\familymc$ a family of MCs and $\phi=\lozenge G$ for $G\subseteq S$
%The \emph{max synthesis problem} is to 
%The problem is defined analogously for an expected reward measure or minimising realisations.
%\end{problem}

\begin{example}[Synthesis problems]
	Recall the family of MCs $\familymc$ from Example~\ref{ex:dtmc_family}.
	For the specification $\varphi=\mathbb{P}_{\geq \nicefrac{1}{10}}(\lozenge \{4\})$, the solution to the threshold synthesis problem is $T=\{r_2,r_4\}$ and $F=\{r_1,r_3\}$, as the goal state $4$ is not reachable for $D_{r_1}$ and $D_{r_3}$.
	For $\phi=\lozenge \{4\}$, the solution to the max synthesis problem on $\familymc$ is $r_2$ or $r_4$, as $D_{r_2}$ and $D_{r_4}$ almost surely reach state $1$.
\end{example}
%
%\begin{theorem}%[NP-completeness]
%		The feasibility synthesis problem is NP-complete~\cite{TACAS}.
%\end{theorem}
%
%The theorem even holds for almost-sure reachability properties. 
%The proof is a straightforward adaption of results for augmented interval Markov chains~\cite[Theorem 3]{DBLP:journals/corr/Chonev17}, partial information games~\cite{DBLP:conf/hybrid/ChatterjeeKS13}, or partially observable MDPs~\cite{DBLP:conf/aaai/ChatterjeeCD16}.

\begin{remark}
It is sometimes beneficial to consider a mild variant of the max-synthesis problem in which the realisation $r^*$ is not required to achieve the maximal reachability probability, but it suffices to be sufficiently close to it.
This notion of \emph{$\varepsilon$-optimal synthesis} for a given $0< \varepsilon \leq 1$ amounts to find a realisation $r^*$ with $\texttt{Prob}(D_{r^*}, \phi) \geq (1{-}\varepsilon) \cdot \max\limits_{r\in \mathcal{R}^{\familymc}} \{\texttt{Prob}(D_{r}, \phi)\}$.
\end{remark}

\subsection{Synthesis Costs}

As in model repair~\cite{bartocci2011model}, it is quite natural to associate non-negative integer costs to the various realisation options.  
This enables distinguishing cheap and expensive repairs.
The realisation (aka: repair) costs should not be confused with the concept of rewards in MCs; the latter impose a cost structure on the MC while realisation costs impose costs on the realisation at hand.

\begin{definition}[Realisation costs]
For family $\familymc$, the function $c:\mathcal{R}^{\familymc} \to \mathbb{N}$ assigns to each realisation $r$ of $\familymc$ a \emph{realisation cost} $c(r)$.
\end{definition}

\noindent
The realisation costs are deliberately defined in a rather abstract manner.
Concrete instances may depend on the probability distribution over $K$, the number of options, weighted combinations thereof, and so forth. 
By imposing an available budget on the possible realisations, we obtain the following cost-dependent variants of the earlier synthesis problems.

\begin{problem}[Cost-constrained synthesis]
Let $\familymc$ be a family of MCs, $\varphi$ and $\phi$ as before, and $B \in \mathbb{N}$ a budget. Consider the synthesis problems:
\begin{enumerate}
\item
Find a realisation $r \in \mathcal{R}^{\familymc}$ with $D_r \models \varphi$ and $c(r) \leq B$.
\item 
Partition $\mathcal{R}^{\familymc}$ with $r \in T$ iff $\left( D_r \models \varphi \mbox{ and } c(r) \leq B \right)$, and $r \in F$ otherwise.
\item
Find $r^* \in \mathcal{R}^{\familymc}$ with  $r^* = \arg\!\!\max\limits_{r\in \mathcal{R}_{\familymc}} \{\texttt{Prob}(D_{r}, \phi) \mid c(r) \leq B \}$.
\end{enumerate}
\end{problem}

\noindent
Cost-constrained maximal synthesis does not need to have a solution; therefore $\arg\!\!\max \emptyset$ equals undefined. 
Cost-optimal versions of the synthesis problems are:

\begin{problem}[Cost-optimal synthesis]
Let $\familymc$ be a family of MCs, $\varphi$ and $\phi$ as before. 
We consider the following cost-optimal synthesis problems:
\begin{enumerate}
\item
Find a realisation $r^* \in T = \{ r \in \mathcal{R}^{\familymc} \mid D_r \models \varphi\}$ with $c(r^*) = \min_{r \in T} \{ c(r) \}$.
\item
Find a minimal-cost realisation $r^*$ for the max/min-synthesis problem.
\end{enumerate}
\end{problem}

\begin{example}[Cost-constrained and cost-optimal synthesis]
Consider our running example, and let the cost of a realisation $r$ be the sum of its number of reachable states and their outgoing transitions.
That is, $c(r_1) = 8, c(r_2) = 10, c(r_3) = 11$, and $c(r_4) = 11$.
For $\phi=\lozenge \{4\}$, and budget $B{=}10$, cost-constrained max synthesis yields $r_2$.
Lowering the budget $B$ to 9, yields $r_1$, while for $B$ less than 8, no realisation is found.
\end{example}

\subsection{A Program Sketching Language}

Probabilistic models are typically specified by means of a high-level modelling language, such as PRISM~\cite{KNP11}, PIOA~\cite{DBLP:journals/tcs/WuSS97}, JANI~\cite{DBLP:conf/tacas/BuddeDHHJT17}, or MODEST~\cite{DBLP:journals/tse/BohnenkampDHK06}. 
Let us briefly describe how the model-based concepts translate to language concepts in  
the PRISM guarded-command language.
The aim is to describe families of MCs, possible constraints on its members, and repair costs in a succinct manner.
A (basic) encoding for the family of Ex.~\ref{ex:dtmc_family} is given in Fig.~\ref{fig:exprism}.
\lstset{language=Prism}   
\newsavebox{\toyprism}
\begin{lrbox}{\toyprism}% Store first listing
\begin{lstlisting}
hole @k2@ either { 2, 3 }
hole @k3@ either { 2, 4 }
module encode
s : [0..4] init 0;
s = 0 -> 0.5: s'=1 + 0.5: s'=@k2@;
s = 1 -> 0.1: s'=0 + 0.9: s'=1;
s = 2 -> 1: s'=@k3@
s = 3 -> 0.2: s'=3 + 0.8: s'=@k3@;
s = 4 -> 1: s'=s
endmodule
\end{lstlisting}
\end{lrbox}
\begin{figure}[t]
\centering
	\usebox{\toyprism}
	\caption{Toy-encoding of the family in Example~\ref{ex:dtmc_family}.}
	\label{fig:exprism}
\end{figure}

A PRISM program consists of one or more reactive modules that may interact with each other. 
Consider a single module.
This is not a restriction as every PRISM program can be flattened into this form. %%  but simplifies the explanation.
A module has a set of (bounded) variables that span its state space.
The possible transitions between states of a module are described by a set of guarded commands of the form:
$$
\mbox{\texttt{guard}} 
\ \ \rightarrow \ \ 
p_1 : \mbox{\texttt{update}}_1 \ldots \ldots + p_n : \mbox{\texttt{update}}_n 
$$
The guard is a boolean expression over the variables of the module. 
If the guard evaluates to true, the module can evolve into a successor state by updating its variables. 
An update is chosen according to the probability distribution given by expressions $p_1$ through $p_n$.
In every state enabling the guard, the evaluation of these expressions must sum up to one.

A PRISM sketch is a program that may contain ``holes''. 
Holes are the unknown parts of the program and can be replaced by one of finitely many options.
A hole is of the form:
$$
\mbox{\texttt{hole} } h \mbox{ \texttt{either} }  
\{ \, \mbox{\texttt{expr}}_1, \ldots, \mbox{\texttt{expr}}_k \, \}
$$
where $h$ is the hole identifier and $\mbox{\texttt{expr}}_i$ is an expression over the program variables.
A hole $h$ can be used in commands in a similar way as a constant, and may occur multiple times within a command. 
%A simple example can be found in Fig.~\ref{fig:pmc_to_family:prism} on page \pageref{fig:pmc_to_family:prism}.
To distinguish cheap and expensive options, options within a hole can have a cost:
$$
\mbox{\texttt{hole} } h \mbox{ \texttt{either} }  
\{ \, x_1 \mbox{\texttt{ is }} \mbox{\texttt{expr}}_1 \mbox{\texttt{ cost }} c_1, 
\ldots, 
x_k \mbox{\texttt{ is }} \mbox{\texttt{expr}}_k \mbox{\texttt{ cost }} c_k
\, \}
$$
where option $i$ is named $x_i$ and has associated cost $c_i$.
Costs can be constants or expressions that evaluate to natural numbers.
The option names $x_1$ through $x_n$ can be used to describe constraint on realisations.
These propositional formulae over option names restrict hole instantiations, e.g.,
$$
\mbox{\texttt{constraint} } (x_1 \, \vee \, x_2) \, \Longrightarrow x_3
$$
requires that whenever the options $x_1$ or $x_2$ are taken for some (potentially different) holes, option $x_3$ (for some hole) is also to be taken. 

The family of realisations of a given PRISM program sketch is now obtained by all possible substitutions of holes $h$ by their options $x_1$ through $x_n$ that satisfy all specified constraints.

\section{Counterexample-Guided Synthesis}
\paragraph{Enumeration.}
A straightforward approach to the synthesis problems for finite families of MCs is to just enumerate all realisations and analyse each of them individually.
This naive method is practically applicable to small- to medium-sized families only.
For more realistic settings, alternative approaches to this baseline are needed.
We present two different counterexample-guided approaches: one based on CEGAR~\cite{DBLP:conf/cav/ClarkeGJLV00} and one on CEGIS~\cite{DBLP:conf/pldi/Solar-LezamaRBE05}.

\begin{figure}
\centering
%\subfigure[]{
%\begin{tikzpicture}[scale=0.35]
%	\draw[step=1.0,black,thin] (0,0) grid (5,5);
%\end{tikzpicture}
%}
\subfigure[]{
\begin{tikzpicture}[scale=0.33]
	\draw[step=1.0,black,thin] (0,0) grid (5,5);
	\draw[fill=blue, opacity=0.6] (0,0) rectangle (5,5);
\end{tikzpicture}
}
\subfigure[]{
\begin{tikzpicture}[scale=0.33]
	\draw[step=1.0,black,thin] (0,0) grid (5,5);
	\draw[fill=red, opacity=0.6] (0,0) rectangle (2,5);
\end{tikzpicture}
}
\subfigure[]{
\begin{tikzpicture}[scale=0.33]
	\draw[step=1.0,black,thin] (0,0) grid (5,5);
	\draw[fill=red, opacity=0.6] (0,0) rectangle (2,5);
	\draw[fill=blue, opacity=0.6] (2,0) rectangle (5,5);
\end{tikzpicture}
}
\subfigure[]{
\begin{tikzpicture}[scale=0.33]
	\draw[step=1.0,black,thin] (0,0) grid (5,5);
	\draw[fill=red, opacity=0.6] (0,0) rectangle (2,5);
	\draw[fill=green, opacity=0.6] (2,0) rectangle (5,2);
\end{tikzpicture}
}\\
%\subfigure[]{
%\begin{tikzpicture}[scale=0.35]
%	\draw[step=1.0,black,thin] (0,0) grid (5,5);
%\end{tikzpicture}
%}
\subfigure[]{
\begin{tikzpicture}[scale=0.33]
	\draw[step=1.0,black,thin] (0,0) grid (5,5);
	\draw[fill=red, opacity=0.6] (0,0) rectangle (1,1);
	\draw[fill=red, opacity=0.4] (0,0) rectangle (1,5);

\end{tikzpicture}
}
\subfigure[]{
\begin{tikzpicture}[scale=0.33]
\draw[step=1.0,black,thin] (0,0) grid (5,5);
\draw[fill=red, opacity=0.6] (0,0) rectangle (1,5);
	\draw[fill=red, opacity=0.6] (1,0) rectangle (2,1);
	\draw[fill=red, opacity=0.4] (1,0) rectangle (2,5);
\end{tikzpicture}
}
\subfigure[]{
\begin{tikzpicture}[scale=0.33]
	\draw[step=1.0,black,thin] (0,0) grid (5,5);
	\draw[fill=red, opacity=0.6] (0,0) rectangle (2,5);
	\draw[fill=green, opacity=0.6] (2,0) rectangle (3,1);
\end{tikzpicture}
}
\subfigure[]{
\begin{tikzpicture}[scale=0.33]
	\draw[step=1.0,black,thin] (0,0) grid (5,5);
	
	\draw[fill=red, opacity=0.6] (0,0) rectangle (2,5);
	\draw[fill=green, opacity=0.6] (2,0) rectangle (3,1);
	
	\draw[fill=green, opacity=0.6] (3,0) rectangle (4,1);
	\draw[fill=green, opacity=0.4] (3,0) rectangle (4,5);
\end{tikzpicture}
}
\vspace{-1em}
\caption{CEGAR (a)--(d) vs. CEGIS (e)--(h) illustrated. The grid depicts a family with two parameters, each with 5 possible values. Thus, each cell corresponds to a realisation. 
Blue indicates a verification call that fails, green/red a covered region satisfying/refuting $\varphi$. The lighter shaded cells indicate realisations ruled out by counterexample analysis.
} 
\label{fig:cegar-and-cegis}
\end{figure}
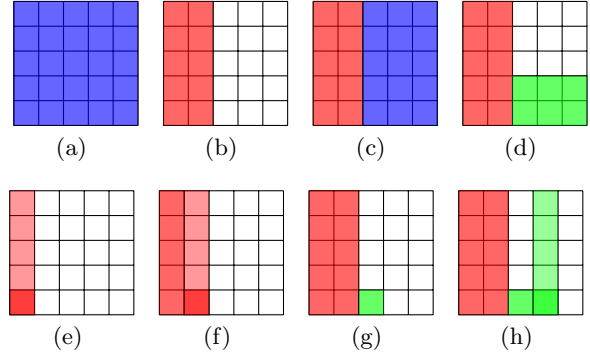

\paragraph{A bird's eye view on our two approaches.}
Let us explain the intuition behind the CEGAR and CEGIS approaches towards synthesis.
Both approaches successively partition the family $\familymc$ into MCs satisfying $\varphi$ and those refuting $\varphi$.
Fig.~\ref{fig:cegar-and-cegis} illustrates this for a two-dimensional parameter space, each parameter having five possible values.
Each cell thus corresponds to a realisation.
CEGAR first checks if all realisations satisfy $\varphi$ on a sound abstraction.
% see Fig.~\ref{fig:cegar-and-cegis}(a).
%If at least one $D_r$ refutes $\varphi$, the verification fails, cf.\ Fig.~\ref{fig:cegar-and-cegis}(b), 
%\sebastian{either due to the subfamily consisting of both satisfying and refuting realisations, or because the abstraction is too coarse. }
Fig.~\ref{fig:cegar-and-cegis}(a) shows the situation when the verification fails, i.e. it gives an inconclusive result. This can happen 
either due to the subfamily consisting of both satisfying and refuting realisations, or because the abstraction is too coarse.
In the next step, CEGAR refines the family into two subfamilies and establishes e.g., that all members in the subfamily represented by the first two columns refute $\varphi$ (indicated in red in Fig.~\ref{fig:cegar-and-cegis}(b)), while verifying the remaining subfamily is again inconclusive (Fig.~\ref{fig:cegar-and-cegis}(c)).
Partitioning that subfamily reveals that the six realisations in the lower two rows fulfil $\varphi$. 
In contrast, CEGIS starts to select a realisation $r$, e.g., the one in the lower left corner.
As $D_r \not\models \varphi$ (indicated dark red in Fig.~\ref{fig:cegar-and-cegis}(e)), the counterexample provided by the verifier rules out all realisations in the leftmost column (indicated in lighter shade).
This scheme is repeated.
In Fig.~\ref{fig:cegar-and-cegis}(f), a realisation is selected (second column, lowest row), and similar to the first case, its counterexample rules out all realisations in that column.
In Fig.~\ref{fig:cegar-and-cegis}(g), a satisfying realisation for $\varphi$ is selected.
In contrast to Fig.~\ref{fig:cegar-and-cegis}(g), Fig.~\ref{fig:cegar-and-cegis}(h) shows that the analysis of counterexamples to $\neg \varphi$ gives rise to more satisfying realisations.
Besides the selected candidate (lowest row), the counterexamples to $\neg \varphi$ cover the entire column.

\subsection{CEGAR}
\label{sec:approach:cegar}
We first represent the family $\familymc$ by a single \emph{all-in-one} MDP.
Selecting action $a_r$ in the (fresh) initial state $s_0^\familymc$ of the MDP corresponds
to choosing the realisation $r \in \mathcal{R}^{\familymc}$ and entering the concrete MC $D_r$.
Let us illustrate this with our example.
\begin{figure}[!h]
	\centering
	\scalebox{0.7}{\begin{tikzpicture}[every node/.style={circle,inner sep=0.1pt}]
	\node[state, initial, initial text=] (s0) {$s_0^\mathfrak{D}$} ;

	\node[draw, right=of s0,yshift=-1cm] (0r1) {$(0,r_2)$};

	\node[draw, right=1.7 cm of 0r1] (1r1) {$(1,r_2)$} ;
	\node[draw, right=1.7 cm of 1r1] (2r1) {$(2,r_2)$};

	\node[draw=gray, right=1.7 cm of 2r1] (3r1) {\color{gray}$(3,r_2)$};
	\node[draw, right=1.7 cm of 3r1] (4r1) {$(4,r_2)$};

	\node[right=0.7 cm of 0r1,fill, circle, inner sep=2pt] (a0r1) {};
	\node[right=0.7 cm of 1r1,fill, circle, inner sep=2pt] (a1r1) {};
	\node[right=0.7 cm of 2r1,fill, circle, inner sep=2pt] (a2r1) {};
	\node[right=0.7 cm of 3r1,fill=gray, circle, inner sep=2pt] (a3r1) {};
	\node[right=0.7 cm of 4r1,fill, circle, inner sep=2pt] (a4r1) {};
	
	\draw[->] (0r1) -- (a0r1);
	\draw[->] (1r1) -- (a1r1);
	\draw[->] (2r1) -- (a2r1);
	\draw[->,gray] (3r1) -- (a3r1);
	\draw[->] (4r1) -- (a4r1);

	\draw[->] (a1r1) edge[bend left=20] node[below] {$0.9$} (1r1);
	\draw[->] (a1r1) edge[bend right=30] node[above] {$0.1$} (0r1);
	\draw[->] (a0r1) edge[] node[above] {$0.5$} (1r1);
	\draw[->] (a0r1) edge[bend right=30] node[pos=0.1,below] {$0.5$} (2r1);
	%\draw[->] (2) edge[loop above] node[auto] {$1$} (2);

	\draw[->] (a4r1) edge[bend left=20] node[auto] {$1$} (4r1);
	%\draw[->,gray] (3) edge[] node[above] {$0.8$} (2);
	%\draw[->,gray] (1) edge[bend right=20] node[pos=0.1,below] {$0.5$} (3);
	
	\draw[->,gray] (a3r1) edge[] node[below] {$0.8$} (4r1);
	\draw[->] (a2r1) edge[bend left=30] node[above, near start] {$1$} (4r1);

	\draw[->,gray] (a3r1) edge[bend left] node[below] {$0.2$} (3r1);
	
	\node[draw, right=of s0,yshift=1cm] (0r2) {$(0,r_1)$};

	\node[draw, right=1.7 cm of 0r2] (1r2) {$(1,r_1)$} ;
	\node[draw, right=1.7 cm of 1r2] (2r2) {$(2,r_1)$};

	\node[draw=gray, right=1.7 cm of 2r2] (3r2) {\color{gray}$(3,r_1)$};
	\node[draw=gray, right=1.7 cm of 3r2] (4r2) {\color{gray}$(4,r_1)$};

	\node[right=0.7 cm of 0r2,fill, circle, inner sep=2pt] (a0r2) {};
	\node[right=0.7 cm of 1r2,fill, circle, inner sep=2pt] (a1r2) {};
	\node[right=0.7 cm of 2r2,fill, circle, inner sep=2pt] (a2r2) {};
	\node[right=0.7 cm of 3r2,fill=gray, circle, inner sep=2pt] (a3r2) {};
	\node[right=0.7 cm of 4r2,fill=gray, circle, inner sep=2pt] (a4r2) {};
	
	\draw[->] (0r2) -- (a0r2);
	\draw[->] (1r2) -- (a1r2);
	\draw[->] (2r2) -- (a2r2);
	\draw[->, gray] (3r2) -- (a3r2);
	\draw[->,gray] (4r2) -- (a4r2);

	\draw[->] (a1r2) edge[bend left=20] node[below] {$0.9$} (1r2);
	\draw[->] (a1r2) edge[bend right=30] node[above] {$0.1$} (0r2);
	
	\draw[->] (a0r2) edge[] node[above] {$0.5$} (1r2);
	\draw[->] (a0r2) edge[bend right=30] node[pos=0.1,below] {$0.5$} (2r2);
	%\draw[->] (2) edge[loop above] node[auto] {$1$} (2);

	\draw[->,gray] (a4r2) edge[bend left=20] node[below] {$1$} (4r2);
	%\draw[->,gray] (3) edge[] node[above] {$0.8$} (2);
	%\draw[->,gray] (1) edge[bend right=20] node[pos=0.1,below] {$0.5$} (3);
	
	\draw[->,gray] (a3r2) edge[bend right=30] node[above] {$0.8$} (2r2);
	\draw[->] (a2r2) edge[bend left=20] node[below, near start] {$1$} (2r2);

	\draw[->,gray] (a3r2) edge[bend left] node[below] {$0.2$} (3r2);
	
	\node[left=0.7 cm of 0r2,fill, circle, inner sep=2pt] (air2) {};
	\node[left=0.7 cm of 0r1,fill, circle, inner sep=2pt] (air1) {};
	
	\draw[->] (s0) -- (air1);
	\draw[->] (s0) -- (air2);
	\draw[->] (air1) edge node[auto] {$1$} (0r1);
	\draw[->] (air2) edge node[auto] {$1$} (0r2);

\end{tikzpicture}}
	\vspace{-0.3cm}	
	\caption{Reachable fragment of the all-in-one MDP $M^\mathfrak{D}$ for realisations $r_1$ and $r_2$.}
	\label{fig:all-in-one}
\end{figure}
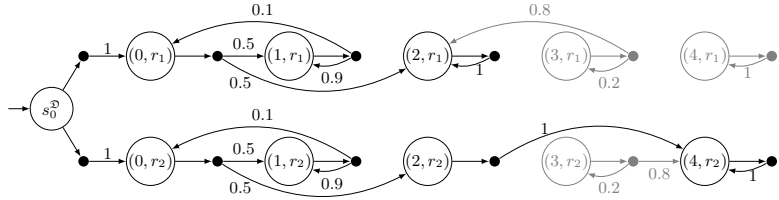
Fig.~\ref{fig:all-in-one} shows the MDP $M^\familymc$ for the family $\familymc$ from Example~\ref{ex:dtmc_family}, where for the sake of readability, only the
transitions and states corresponding to realisations $r_1$ and $r_2$ are included.
Transitions to states $(s, r_i)$ are labeled with action $a_{r_i}$; these action labels are omitted here.
Unreachable states from the initial MDP state $s_0^\familymc$ are marked grey. 
There is a one-to-one relationship between a deterministic memoryless scheduler of MDP $M^\familymc$ and a realisation of $\familymc$.
Thus, model checking $M^\familymc$ yields extremal probabilities for all realisations of the family $\familymc$.

\begin{figure}[h]
	\centering
	\scalebox{0.7}{\begin{tikzpicture}[every node/.style={circle,inner sep=0.1pt}]
	\node[state, initial, initial text=] (s0) {$s_0^\mathfrak{D}$} ;

	\node[draw, right=of s0] (0r1) {$[0]_{\sim}$};

	\node[draw, right=1.7 cm of 0r1] (1r1) {$[1]_{\sim}$} ;
	\node[draw, right=1.7 cm of 1r1] (2r1) {$[2]_{\sim}$};

	\node[draw, right=1.7 cm of 2r1] (3r1) {$[3]_{\sim}$};
	\node[draw, right=1.7 cm of 3r1] (4r1) {$[4]_{\sim}$};

	\node[below=0.7 cm of 0r1,fill, circle, inner sep=2pt] (a0r1) {};
	\node[below=0.7 cm of 1r1,fill, circle, inner sep=2pt] (a1r1) {};
	\node[below=0.7 cm of 2r1,fill, circle, inner sep=2pt] (a2r1) {};
	\node[below=0.7 cm of 3r1,fill, circle, inner sep=2pt] (a3r1) {};
	\node[below=0.7 cm of 4r1,fill, circle, inner sep=2pt] (a4r1) {};
	
	\draw[->] (0r1) -- (a0r1);
	\draw[->] (1r1) -- (a1r1);
	\draw[->] (2r1) -- (a2r1);
	\draw[->] (3r1) -- (a3r1);
	\draw[->] (4r1) -- (a4r1);

	\draw[->] (a1r1) edge[bend right=30] node[right] {$0.9$} (1r1);
	\draw[->] (a1r1) edge[] node[above,pos=0.15] {$0.1$} (0r1);
	\draw[->] (a0r1) edge[] node[above,pos=0.15] {$0.5$} (1r1);
	\draw[->] (a0r1) edge[bend right=30] node[pos=0.1,below] {$0.5$} (2r1);
	%\draw[->] (2) edge[loop above] node[auto] {$1$} (2);

	\draw[->] (a4r1) edge[bend right=30] node[right] {$1$} (4r1);
	%\draw[->,gray] (3) edge[] node[above] {$0.8$} (2);
	%\draw[->,gray] (1) edge[bend right=20] node[pos=0.1,below] {$0.5$} (3);
	
	\draw[->] (a3r1) edge[] node[above,pos=0.6] {$0.8$} (4r1);
	\draw[->] (a2r1) edge[bend right=30] node[above, near start] {$1$} (4r1);

	\draw[->] (a3r1) edge[bend right=30] node[right] {$0.2$} (3r1);

	\node[above=0.7 cm of 0r1,fill, circle, inner sep=2pt] (a0r2) {};
	\node[above=0.7 cm of 1r1,fill, circle, inner sep=2pt] (a1r2) {};
	\node[above=0.7 cm of 2r1,fill, circle, inner sep=2pt] (a2r2) {};
	\node[above=0.7 cm of 3r1,fill, circle, inner sep=2pt] (a3r2) {};
	\node[above=0.7 cm of 4r1,fill, circle, inner sep=2pt] (a4r2) {};
	
	\draw[->] (0r1) -- (a0r2);
	\draw[->] (1r1) -- (a1r2);
	\draw[->] (2r1) -- (a2r2);
	\draw[->] (3r1) -- (a3r2);
	\draw[->] (4r1) -- (a4r2);

	\draw[->] (a1r2) edge[bend left=30] node[right] {$0.9$} (1r1);
	\draw[->] (a1r2) edge[] node[below,pos=0.15] {$0.1$} (0r1);
	
	\draw[->] (a0r2) edge[] node[below,pos=0.15] {$0.5$} (1r1);
	\draw[->] (a0r2) edge[bend left=30] node[pos=0.1,above] {$0.5$} (2r1);
	%\draw[->] (2) edge[loop above] node[auto] {$1$} (2);

	\draw[->] (a4r2) edge[bend left=30] node[right] {$1$} (4r1);
	%\draw[->,gray] (3) edge[] node[above] {$0.8$} (2);
	%\draw[->,gray] (1) edge[bend right=20] node[pos=0.1,below] {$0.5$} (3);
	
	\draw[->] (a3r2) edge[] node[above] {$0.8$} (2r1);
	\draw[->] (a2r2) edge[bend left=30] node[right, near start] {$1$} (2r1);

	\draw[->] (a3r2) edge[bend left] node[right] {$0.2$} (3r1);
	
	\node[left=0.7 cm of 0r1,fill, circle, inner sep=2pt] (air1) {};
	
	\draw[->] (s0) -- (air1);
	\draw[->] (air1) edge node[auto] {$1$} (0r1);

\end{tikzpicture}}\vspace{-0.3cm}
	\caption{The quotient MDP $M^\mathfrak{D}_\sim$ for realisations $r_1$ (top actions) and $r_2$ (bottom actions)}
	\label{fig:forgotten}
\end{figure}
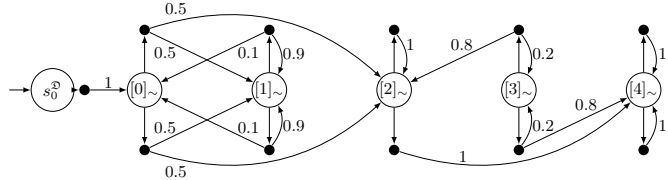

The MDP model grows linearly with the number of family members. To mitigate the complexity, we apply a simple abstraction where 
the realisation of a state in the MDP $M^\familymc$ is abstracted away, i.e. the item $r$ is ignored in state $(s,r)$.
Applying this to our running example amounts to a column-wise grouping of states in Fig.~\ref{fig:all-in-one}.
This results in the quotient MDP $M^\mathfrak{D}_\sim$ in Fig.~\ref{fig:forgotten}.
By the over-approximation in the quotient MDP $M^\mathfrak{D}_\sim$, a scheduler may first choose actions $a_r$ and then $a_{r'}$.
This corresponds to switching from realisation $r$ to $r'$.
Such inconsistent regimes result in an MC outside the family $\familymc$.
There is one-to-one relationship between consistent schedulers---those that globally stick to a single realisation $r$---and the realisation of $\familymc$.
%Thus, a consistent deterministic scheduler on the quotient $M^\mathfrak{D}_\sim$ corresponds one-to-one to a deterministic memoryless scheduler on MDP $M^\familymc$.
%%
\begin{example}
Consider the quotient MDP $M^\mathfrak{D}_\sim$ in Fig.~\ref{fig:forgotten}.
Transitions from previously unreachable states, marked grey before, are reachable in the quotient. 
The scheduler $\sched$ on the quotient $M^\mathfrak{D}_\sim$ that picks $a_{r_1}$ in $[2]_\sim$ and $a_{r_2}$ in state $[3]_\sim$ is inconsistent, as it behaves according to two different realisations $r_1$ and $r_2$.
\end{example}

\subsection{CEGIS}
\begin{figure}[t]
\centering
\subfigure[Fragment of $r_1$]{
\scalebox{0.7}{\begin{tikzpicture}[every node/.style={circle}]
	\node[draw] (1) {$0$};
	\node[draw, right=1.2 cm of 1] (0) {$1$} ;
	\node[draw, right=1.2 cm of 0] (2) {$2$};
	%\node[draw=gray, right=1.2 cm of 2] (3) {\color{gray}$3$};
	%\node[draw=gray, right=1.2 cm of 3] (4) {\color{gray}$4$};

	%\draw[->] (0) edge[loop above] node[auto] {$0.9$} (0);
	%\draw[->] (0) edge[bend right=60] node[above] {$0.1$} (1);
	\draw[->] (1) edge[] node[above] {$0.5$} (0);
	\draw[->] (1) edge[bend right=20] node[pos=0.1,below] {$0.5$} (2);
	%\draw[->] (2) edge[loop above] node[auto] {$1$} (2);

	%\draw[->,gray] (4) edge[loop above] node[auto] {$1$} (4);
	%\draw[->,gray] (3) edge[] node[above] {$0.8$} (2);
	%\draw[->,gray] (1) edge[bend right=20] node[pos=0.1,below] {$0.5$} (3);
	
	%\draw[->,gray] (3) edge[] node[below] {$0.8$} (4);
	%\draw[->,gray] (2) edge[bend left=20] node[above, near start] {$1$} (4);

	%\draw[->,gray] (3) edge[loop below] node[left] {$0.2$} (3);

%	\draw[->,gray] (1) edge[] node[below] {$0.5$} (2);
%	\draw[->] (0) edge[loop below] node[auto] {$0.5$} (0);
%	\draw[->,gray] (2) edge[loop below] node[auto] {$1$} (2);
%	\draw[->,gray] (3) edge[bend right=20] node[above, near start] {$0.5$} (0);
%	\draw[->,gray] (3) edge[] node[below] {$0.5$} (2);
	\draw ($(1.north) + (0,0.3)$) edge[->] (1);
	\draw  [use as bounding box, draw=white] (-1.2,-0.4) rectangle (6.1,0.8) {};%different BB to other pics because of weird spacing
%different BB to other pics because of weird spacing
\end{tikzpicture}%
}
	\label{fig:fragment}
}
	\subfigure[Sub-MC of $D_{r_1}$ with $C=\{0\}$]{
	\scalebox{0.7}{\begin{tikzpicture}[every node/.style={circle}]
	\node[draw] (1) {$0$};
	\node[draw, right=1.2 cm of 1] (0) {$1$} ;
	\node[draw, right=1.2 cm of 0] (2) {$2$};
	%\node[draw=gray, right=1.2 cm of 2] (3) {\color{gray}$3$};
	%\node[draw=gray, right=1.2 cm of 3] (4) {\color{gray}$4$};

	\draw[->] (0) edge[loop right] node[auto] {$1$} (0);
	\draw[->] (2) edge[loop right] node[auto] {$1$} (2);
	
	%\draw[->] (0) edge[bend right=60] node[above] {$0.1$} (1);
	\draw[->] (1) edge[] node[above] {$0.5$} (0);
	\draw[->] (1) edge[bend right=20] node[pos=0.1,below] {$0.5$} (2);
	
	%\draw[->] (2) edge[loop above] node[auto] {$1$} (2);

	%\draw[->,gray] (4) edge[loop above] node[auto] {$1$} (4);
	%\draw[->,gray] (3) edge[] node[above] {$0.8$} (2);
	%\draw[->,gray] (1) edge[bend right=20] node[pos=0.1,below] {$0.5$} (3);
	
	%\draw[->,gray] (3) edge[] node[below] {$0.8$} (4);
	%\draw[->,gray] (2) edge[bend left=20] node[above, near start] {$1$} (4);

	%\draw[->,gray] (3) edge[loop below] node[left] {$0.2$} (3);

%	\draw[->,gray] (1) edge[] node[below] {$0.5$} (2);
%	\draw[->] (0) edge[loop below] node[auto] {$0.5$} (0);
%	\draw[->,gray] (2) edge[loop below] node[auto] {$1$} (2);
%	\draw[->,gray] (3) edge[bend right=20] node[above, near start] {$0.5$} (0);
%	\draw[->,gray] (3) edge[] node[below] {$0.5$} (2);
	\draw ($(1.north) + (0,0.3)$) edge[->] (1);
	\draw  [use as bounding box, draw=white] (-1.2,-0.4) rectangle (6.1,0.8) {};%different BB to other pics because of weird spacing
%different BB to other pics because of weird spacing
\end{tikzpicture}%
}
	\label{fig:submodel}
	}
	\caption{Fragment and corresponding sub-MC that suffices to refute $\varphi$}
\end{figure}
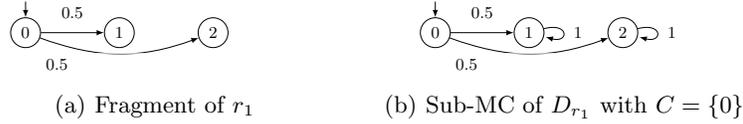

We follow the typical separation of concerns as in oracle-guided inductive synthesis~\cite{DBLP:journals/cacm/AlurSFS18,DBLP:conf/kbse/GerasimouTC15,DBLP:journals/ftpl/GulwaniPS17}:
a \emph{synthesiser} selects single realisations that have not been considered before, and a \emph{verifier} checks the selected realisation.
Let us first focus on the \emph{verifier}.
Consider our running example with $\varphi = \mathbb{P}_{\leq \nicefrac{2}{5}}(\lozenge \{ 2 \})$.
Assume the synthesiser picks realisation $r_1$.
The verifier then builds $D_{r_1}$ and determines $D_{r_1} \not\models \varphi$.
Observe that the verifier does not need the full realisation $D_{r_1}$ to refute $\varphi$.
In fact, the paths in the fragment of $D_{r_1}$ in Fig.~\ref{fig:fragment} 
%% (ignoring the outgoing transitions of states $1$ and $2$) 
suffice to show that the probability to reach state $2$ exceeds $\nicefrac 2 5$.
Formally, the fragment in Fig.~\ref{fig:submodel} is a sub-MC with critical states $C = \{ 0 \}$.
The essential property is~\cite{DBLP:journals/tcs/WimmerJAKB14}:
\[
\mbox{\emph{If a sub-MC of a MC $D$ refutes the safety property $\varphi$, then $D$ refutes $\varphi$ too.}}
\]
Now observe that the considered sub-MC is part of realisation $r_2$ too. 
Thus, $D_{r_2} \not\models \varphi$.
This can be concluded by considering $r_2$, $\familymc$ and $C$ \emph{without constructing $D_{r_2}$}.
First, take the parameters occurring in $\mathfrak{P}(c)$ for any $c \in C$.
This yields $k_1$ and $k_2$. 
The values for the other parameters thus do not affect the shape of the sub-MC induced by $C$.
Realisation $r_2$ only varies from $r_1$ in the value of $k_3$.
Therefore, the sub-MC of $D_{r_2}$ induced by $C$ is isomorphic to the sub-MC of $D_{r_1}$ induced by $C$.
This results in concluding $D_{r_2} \not\models \varphi$. 
%% \mcc{It is hard to parse this paragraph, however, I am do not how to fix it without extending the text.}

Let us generalise our observations:
The verifier gets a realisation $r$, builds $D_r$ and checks whether $D_r \models \varphi$. 
If yes, then $\neg\varphi$ is considered to seek for other realisations satisfying $\varphi$\footnote{Note that $\neg\varphi$ is not a safety property, but the presented idea can be extended to liveness properties too.}. 
Otherwise, some sub-MC of $D_r$ refutes $\varphi$.
(These counterexamples can be constructed from MC models using techniques from~\cite{DBLP:journals/tcs/WimmerJAKB14}, or as snippets of PRISM programs using~\cite{DBLP:conf/atva/DehnertJWAK14}.)
The verifier constructs the conflict set $C$ and checks which parameters $K'$ occur on the outgoing transitions of states in $C$.
Each $r'$ with the same parameter values for $K'$ can be immediately refuted, without building the realisation $D_{r'}$.
Put in a nutshell, counterexamples are exploited to rule out several realisations in one shot.

The main task that remains for the \emph{synthesiser} is to book-keep the considered and excluded realisations, and, heuristically select realisations that lead to small\footnote{For some suitable measure of size.} counterexamples, as these exclude potentially many realisations.

%The sets of critical states are not unique: by adapting ideas from high-level counterexamples, we can prefer such sets that allow to exclude many realisations.

%%\subsection{CEGAR vs CEGIS}
%%{\color{red} Short recap: Advantages of either method}
%%
%%We emphasise that the nature of the counterexamples differs for the two approaches.
%%
%%CEGAR downsides:
%%- quotient may be significantly larger
%%- hard to incorporate constraints on the family
%%
%%An open challenge is to develop effective combinations of these orthogonal methods.

\section{Applications}
\label{sec:applications}
This section illustrates the potential wide applicability of probabilistic model synthesis by providing examples from three different areas: program sketching, software product lines, and controller synthesis for partially observable models.

\subsection{Program Sketching}

\paragraph{Background.}
The idea of program sketching~\cite{DBLP:journals/sttt/Solar-Lezama13} is to start with a program sketch, a partial program in which difficult expressions, guards, and statements are left unspecified.
The hypothesis of program sketching is that programmers often have an idea about the main control flow of the program but filling in all low-level details is laborious and error prone.
Completing these low-level details is left to an automated synthesiser.
Syntax generators are used to describe a space of the possible code fragments that can be used to complete the program.
The synthesised program has to satisfy the specification $\varphi$.
Program sketching has been successfully applied to e.g., scientific programs and concurrent data structures~\cite{DBLP:journals/cacm/AlurSFS18}.
We show how our approaches can be applied to sketching of probabilistic programs. 

\paragraph{Concrete challenge.}
Our program sketch is describing a \emph{dynamic power manager (DPM)}, a key component in dynamically optimising energy consumption~\cite{DBLP:journals/tcad/BeniniBPM99}.
A DPM controls changing the system's power states at run time. 
Depending on workload and performance constraints, it issues commands (e.g., go into sleep mode, wake up) to the system.
We consider a DPM system with two request priorities, low and high; these priorities typically depend on the time-criticality. 
Requests are placed in finite buffers (based on their priority), provided the buffers are not full. 
Otherwise, the requests are lost.
A similar DPM has been analysed by probabilistic model checking~\cite{DBLP:journals/tcad/SesicDM08}.

\paragraph{Problem statement.}
Our goal is to \emph{synthesise a DPM program} that decides to switch power state based on the current workload expressed in terms of the occupancy of the low-priority and high-priority request buffers. 

\paragraph*{Approach.}
The starting point is a program sketch that includes partially specified commands of the form:
\[
g_H \ \ \& \ \ g_L \quad \longrightarrow \quad 1: \mbox{\sl state}' = X
\]
where $g_H$ and $g_L$ are partially specified guards concerning the low-priority and high-priority request buffer, respectively, and hole $X$ represents an unknown update of the DPM state.
%% In particular, $g_H$ ($g_L$, respectively) expresses whether the occupancy of the buffer is within an underspecified interval (e.g., between 0 and half of the queue capacity) and the DPM state controls the transition of the energy-critical component (e.g., if \mbox{\sl state} = 1, the DPM sends a control signal to switch to an active state). 
%%
Possible code fragments to complete the commands are e.g., in \mbox{\sl state} = 1, the DPM sends a control signal to switch to an active state,  while $g_H$ (and similarly $g_L$) indicates that the occupancy is within a given interval, e.g., at most 50\%.
We synthesise the guards and updates such that the resulting DPM control program meets a conjunction of objectives (inspired by~\cite{DBLP:conf/kbse/GerasimouTC15}) that constrain the expected number of lost low- and high-priority requests and the expected energy consumption, for different thresholds $\lambda$ imposed on these expected values.

\paragraph{Results.}
The MC family has over $3\cdot10^5$ realisations, i.e., control programs. The average realisation has more than 5000 states. We consider an unsatisfiable conjunction of 3 properties describing a possible DPM specification.
Within 20 minutes, the conjunction is shown to be unsatisfiable (although each property alone is satisfiable). 
An enumerative approach takes more than 20 hours to show this.
For a satisfiable conjunction, we find a realisation within minutes.

\subsection{Software Product Lines}

\paragraph{Background.}
A software product line is (according to wikipedia) ``a set of software-intensive systems that share a common, managed set of features satisfying the specific needs of a particular market segment or mission and that are developed from a common set of core assets in a prescribed way''.
Products in a software product line have different features which can be understood as functionalities changing the behaviours of a core software system.
They thus provide an elegant way to specify families of systems: every member of the family comprises the core system and a combination of features.
Randomness appears when modelling energy consumption or failure probabilities. 

\paragraph{Concrete challenge.}
We consider the BSN (Body Sensor Network) software product line benchmark from~\cite{RodriguesANLCSS15}, the largest benchmark analysed by probabilistic model checking with the ProFeat tool~\cite{DBLP:journals/fac/ChrszonDKB18}.
BSN describes a network of connected sensors that send measurements to a unit identifying health-critical situations. 
The family contains the various configurations of 10 binary features, that is, whether a sensor is available or not.
We are interested in the reliability of the system, that is, in the   probability that the system behaves as described.
The system is described as a parametric Markov chain:
\lstset{language=Prism}   
\newsavebox{\featuresencoded}
\begin{lrbox}{\featuresencoded}% Store first listing
\begin{lstlisting}
hole @fa@ either { 0, 1 }
hole @fb@ either { 0, 1 }
module encode
s : [0..1] init 0;
FA : [0..1] init 0;
FB : [0..1] init 0;
s = 0 -> 1: s'=1 & FA'=@fa@ & FB'=@fb@;
..
endmodule
\end{lstlisting}
\end{lrbox}

\begin{figure}[t]
\centering
	\subfigure[Parametric MC]{
	\scalebox{0.7}{
	\begin{tikzpicture}[every node/.style={circle}]
		\node[draw] (s) {$s$}; 
		\node[right=of s] (dummy) {};
		\node[draw, above=0.35cm of dummy] (s10) {\scriptsize$10$};
		\node[draw, above=0.6cm of s10] (s11) {\scriptsize$11$};
		\node[draw, below=0.35cm of dummy] (s01) {\scriptsize$01$};
		\node[draw, below=0.6cm of s01] (s00) {\scriptsize$00$};
		
		\draw[->] (s) edge[bend right=15] node [pos=0.7,right] {$(1{-}f_a){\cdot}(1{-}f_b)$} (s00);
		\draw[->] (s) edge[bend left=15] node [pos=0.1, right] {$f_a{\cdot}(1{-}f_b)$} (s10);
		\draw[->] (s) edge[bend right=15] node [pos=0.1, right] {$(1{-}f_a){\cdot}f_b$} (s01);
		\draw[->] (s) edge[bend left=15] node [pos=0.7,right] {$f_a{\cdot}f_b$} (s11);
	\end{tikzpicture}
	}
	\label{fig:pmc_to_family:pmc}
	}
	\subfigure[The PRISM encoding for the family]{
	\raisebox{2cm}{\usebox{\featuresencoded}}
	\label{fig:pmc_to_family:prism}
	}
	\caption{Translating a parametric MC formulation to an encoding of an family of DTMCs.}
\end{figure}
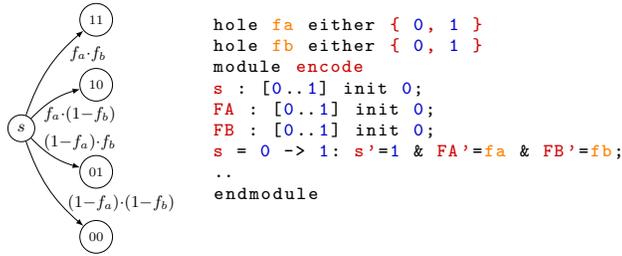

\begin{example}
We consider a variation point (a state whose future behaviour depends on the features) where depending on the availability of features $F_a, F_b$ the model behaves differently. 
For each feature, a Boolean parameter $f$ is 1 if the feature is active and 0 otherwise. 
At a variation point, the probability of every transition is scaled by factor $p$, which equals $f$ if the feature enables the transition and $1{-}f$ otherwise.
This results in parametric MC in Fig.~\ref{fig:pmc_to_family:pmc}.
\end{example}

\paragraph{Problem statement.}
Find all features combinations where the induced system does not meet a certain reliability.

\paragraph{Approach.}
The formulation in~\cite{RodriguesANLCSS15} is a parametric MC, and therefore seems amenable to standard parameter synthesis in which probabilities have to be synthesised. 
However, in absence of certain features, transitions are taken with probability zero. 
Traditional parameter synthesis techniques do not allow for such assignments.
We translate the parametric MC into a PRISM-description of a family, as illustrated in the following example:
\begin{example}
	 We adapt the encoding in Fig.~\ref{fig:pmc_to_family:pmc} to an encoding in Fig.~\ref{fig:pmc_to_family:prism}.
\end{example}

\paragraph{Results.}
Though this is the largest product line example used in~\cite{DBLP:journals/fac/ChrszonDKB18}, verifying a family with just 1024 family members, with an average size of the realisation of roughly 100 states is mostly trivial.
Within seconds, we can categorise the different realisations based on their reliability, either by our approaches or by enumeration. 

\subsection{Controller Synthesis in Partially Observable Systems}

\paragraph{Background.}
As a next application, we consider controller synthesis (aka: scheduler synthesis) in partially-observable MDPs (POMDPs, for short).
A POMDP~\cite{DBLP:journals/ai/KaelblingLC98} is an MDP in which an observation $o(s)$ is associated with every state $s$.
POMDP controllers do not have access to the current state of the POMDP; instead, they can only use the observations of the visited states.
Thus, whereas an MDP scheduler bases its decisions on finite paths of the form $\pi = s_0 \xrightarrow{\act_0} \cdots \xrightarrow{\act_{n-1}} s_n$, a POMDP controller does so using the observation sequence $o(\pi) = o(s_0) \xrightarrow{\act_0} \cdots \xrightarrow{\act_{n-1}} o(s_n)$.
Several paths in the underlying MDP $M$ may give rise to the same observation sequence. 
Controllers have to take this restricted observability into account: They cannot distinguish paths with the same observation sequence.
Controller synthesis for POMDPs is notoriously hard: Finding an optimal strategy is
undecidable~\cite{DBLP:journals/jcss/ChatterjeeCT16} and finding an optimal memoryless strategy is already NP- and SQRT-SUM hard~\cite{DBLP:journals/toct/VlassisLB12}. 
The complexity of the problem makes the possibility to guide the search for a strategy by means of synthesis very interesting. 

\paragraph{Concrete challenge.}
We consider Maze, a classical motion planning problem considered as POMDP, see e.g., \cite{DBLP:conf/uai/MeuleauKKC99}.
A robot is put in a maze with paths surrounded by walls, and its aim is to go to a goal position in the maze.
The problem is partially observ­able because the robot cannot perceive its true location, but only the presence or the absence of a wall on either side of its current position.  
There is a non-zero prob­ability of slipping, so that the robot does not always know if its last attempt to make a move had any consequence on its actual position in the maze.

\paragraph{Problem statement.}
The objective is to \emph{synthesise a deterministic finite-state controller}~\cite{DBLP:conf/uai/MeuleauKKC99,DBLP:conf/aaai/ChatterjeeCD16} for a Maze (of different size) with a bounded number of states that minimises the expected time for the robot to reach the goal.

\paragraph{Approach.}
To cast this POMDP problem in our framework of families of MCs, we adapt a recent result~\cite{DBLP:conf/uai/Junges0WQWK018} that established a one-to-one correspondence between finding finite memory randomised controllers in POMDPs and satisfying parameter valuations in parametric MCs.
We sketch a controller by restricting the memory to a fixed bound. 
Costs are used to penalise the complexity of the controllers such that simple, i.e., easy implementable, finite-memory controllers result.
The family describes all MCs induced by small-memory observation-based deterministic strategies with a fixed upper bound on their amount of memory. 
We are interested in the expected time to the goal.
(This problem can be formalised by adding rewards to MDPs in the usual way.)

\paragraph{Results.}
The MC family has a bit more than $10^6$ realisations, i.e., observation-based strategies. The average realisation has 134 states.
Among other results, within seconds, we find the 4 strategies that were at most 2\% off the maximum. 
In comparison, an enumeration-based approach takes several hours, and enumerating all consistent strategies of the quotient (see Sect.~\ref{sec:approach:cegar}) takes more than an hour.

\section{Epilogue}

\paragraph{Summary.}
This paper outlined two techniques for the automated synthesis of finite-state probabilistic models or programs.
The CEGAR approach takes as \mc{ a starting point} an abstract representation of a family of Markov chains and exploits inconsistent policies --- policies that switch between different realisations --- to iteratively refine the design space.
The CEGIS approach exploits critical subsystems as counterexamples and uses SMT techniques to prune the design space by analysing the counterexamples.
We foresee a wide applicability of these kind of synthesis techniques; we illustrated this by examples from program sketching, controller synthesis, and software product lines.
Both techniques significantly outperform a naive enumerative approach and differ substantially from the few existing approaches to synthesising probabilistic programs~\cite{DBLP:conf/pldi/NoriORV15, DBLP:conf/cav/ChasinsP17}.
CEGAR works particularly well if the quotient MDP is succinct while CEGIS excels the ``more unsatisfiable'' the synthesis problem is.
CEGAR has difficulties treating constraints on family members (which are straightforward with CEGIS), whereas the performance of CEGIS significantly drops for synthesis problems for which the threshold is close to the true reachability probability. 

%%
%%Data-driven synthesis of probabilistic proigrams~\cite{DBLP:conf/cav/ChasinsP17}
%%we have created a tool that synthesizes PPL programs from relational datasets. Our synthesizer leverages the input data to generate a program sketch, then applies simulated annealing to complete the sketch. We introduce a data-guided approach to the program mutation stage of simulated annealing; 
%%
%%\cite{DBLP:conf/pldi/NoriORV15} presents an efficient Markov Chain Monte Carlo (MCMC) based synthesis algorithm to instantiate the holes in the sketch with program fragments. 

\paragraph{Future work.}
The approaches in this paper are first stepping stones towards the automated synthesis of probabilistic models.
This topic has plenty of interesting directions for future work.
This includes synthesising infinite-state probabilistic programs, integrating efficient parameter synthesis and model synthesis, developing adequate modeling formalisms for families of probabilistic models, and learning algorithms for probabilistic models~\cite{DBLP:journals/ml/MaoCJNLN16,DBLP:journals/sttt/WangSYP18}.

{\small
\paragraph{Acknowledgement.}
This chapter is a birthday salute to Scott Smolka on the occasion of his 65th birthday.
Scott's research is extremely novel --- he is always ``ahead of the pack''.
He pioneered probabilistic aspects in formal modeling and verification with his seminal works on probabilistic processes, testing pre-orders, and approximate bisimulation.
His work with Grosu on Monte Carlo model checking emerged into (what others misnamed) statistical model checking.
Scott was the first to combine logical programming with model checking and applied formal methods to new applications such as cardiac devices and, more recently, bird flocking.
His work has been (and still is) an enormous source of inspiration.
This paper celebrates his (and his co-authors') work on model repair for probabilistic models and illustrates how tweaking probabilities (as in model repair) can be generalised towards synthesising model structures.
Happy birthday, Scott!
}

\bibliographystyle{splncs04}
\bibliography{literature}

%%\clearpage
%%\appendix
%%\section*{Appendix}
\end{document}